\documentclass{article}
\usepackage{cite}
\usepackage{graphicx}
\usepackage{dcolumn}

\begin{document}

\date{}
\title{On the Rayleigh-Ritz variational method}
\author{Francisco M. Fern\'{a}ndez \thanks{%
E-mail: fernande@quimica.unlp.edu.ar} \\
%EndAName
INIFTA, Divisi\'on Qu\'imica Te\'orica,\\
Blvd. 113 S/N, Sucursal 4, Casilla de Correo 16,\\
1900 La Plata, Argentina.}
\maketitle

\begin{abstract}
We give a simple proof of the well known fact that the approximate
eigenvalues provided by the Rayleigh-Ritz variational method are
increasingly accurate upper bounds to the exact ones. To this end, we resort
to the variational principle, mentioned in most textbooks on quantum
chemistry, and to a well known set of projection operators. We think that
present approach may be suitable for an advanced course on quantum mechanics
or quantum chemistry.
\end{abstract}

\section{Introduction}

\label{sed:intro}

The Rayleigh-Ritz variational method (RRVM) is discussed in most textbooks
on quantum chemistry\cite{P68,SO96} but it is not so widely found in
textbooks on quantum mechanics\cite{CDL77}. The reason is that this approach
is most important for the study of the electronic structure of molecules\cite%
{P68,SO96}. The RRVM provides increasingly tighter upper bounds to all the
eigenvalues of an Hermitian operator as proved long ago by MacDonald\cite%
{M33}. This property of the approach is invoked in many textbooks without
proof\cite{P68}. Szabo and Ostlund\cite{SO96} sketched an incomplete simple
proof for the two lowest eigenvalues in Exercise 1.21 of page 36. In this
paper we generalize this result to all the eigenvalues of a given Hermitian
operator in a way that may be suitable for a graduate course on quantum
mechanics or quantum chemistry.

In section~\ref{sec:Var_princ} we outline the variational principle already
mentioned in most textbooks\cite{P68,SO96,CDL77}. In section~\ref%
{sec:projection} we introduce some well known projection operators\cite%
{P68,SO96,CDL77} that facilitate the derivation of the proof of the main
property of the RRVM in section~\ref{sec:RR}. In section~\ref{sec:example}
we apply the RRVM to a simple nontrivial example. Finally, in section~\ref%
{sec:conclusions} we summarize the main results and draw conclusions.

\section{Variational principle}

\label{sec:Var_princ}

We are interested in the eigenvalue equation
\begin{eqnarray}
H\psi _{n} &=&E_{n}\psi _{n},\;n=1,2,\ldots ,  \nonumber \\
E_{1} &\leq &E_{2}\leq \ldots ,\;\left\langle \psi _{i}\right. \left| \psi
_{j}\right\rangle =\delta _{ij},  \label{eq:eigenvalue_eq_H}
\end{eqnarray}
for a Hermitian operator $H$. If $\varphi $ belongs to the domain of $H$
then
\begin{equation}
\varphi =\sum_{j}c_{j}\psi _{j},\;c_{j}=\left\langle \psi _{j}\right. \left|
\varphi \right\rangle ,  \label{eq:trial_fun}
\end{equation}
and
\begin{eqnarray}
\left\langle \varphi \right| H\left| \varphi \right\rangle &=&\sum_{j}\left|
c_{j}\right| ^{2}E_{j}\geq E_{1}\sum_{j}\left| c_{j}\right|
^{2}=E_{1}\left\langle \varphi \right. \left| \varphi \right\rangle
\Rightarrow  \nonumber \\
\frac{\left\langle \varphi \right| H\left| \varphi \right\rangle }{%
\left\langle \varphi \right. \left| \varphi \right\rangle } &\geq &E_{1}.
\label{eq:UB_E_1}
\end{eqnarray}

If
\begin{equation}
\left\langle \psi _{j}\right. \left| \varphi \right\rangle =0,\;j=1,2,\ldots
,k-1,  \label{eq:orthogonality_condition}
\end{equation}
then the same argument leads to
\begin{equation}
\frac{\left\langle \varphi \right| H\left| \varphi \right\rangle }{%
\left\langle \varphi \right. \left| \varphi \right\rangle }\geq E_{k}.
\label{eq:UB_E_k}
\end{equation}
In general, this result may not be of practical utility because it requires
some of the supposedly unknown eigenvectors $\psi _{j}$ of $H$. However, if
we know the symmetry of such eigenvectors we can construct a trial function $%
\varphi $ that satisfies the orthogonality condition (\ref%
{eq:orthogonality_condition}) for some value of $k$. We will resort to
equations (\ref{eq:UB_E_1}-\ref{eq:UB_E_k}) to derive the main results of
this paper in section~\ref{sec:RR}.

\section{Projection operators}

\label{sec:projection}

By means of an orthonormal basis set $\left\{ \left| j\right\rangle
,\;j=1,2,\ldots \right\} $, $\left\langle i\right. \left| j\right\rangle
=\delta _{ij}$, we can construct projection operators of the form\cite%
{P68,SO96,CDL77}
\begin{equation}
P_{N}=\sum_{j=1}^{N}\left| j\right\rangle \left\langle j\right|
,\;N=1,2,\ldots ,  \label{eq:P_N}
\end{equation}
that are Hermitian ($P_{N}^{\dagger }=P_{N}$) and idempotent
\begin{equation}
P_{N}^{2}=\sum_{i=1}^{N}\sum_{j=1}^{N}\left| i\right\rangle \left\langle
i\right| \left| j\right\rangle \left\langle j\right| =\sum_{j=1}^{N}\left|
j\right\rangle \left\langle j\right| =P_{N}.  \label{eq:P_N^2}
\end{equation}
Since $P_{N}=P_{N-1}+\left| N\right\rangle \left\langle N\right| $ and $%
\left| N\right\rangle \left\langle N\right| P_{N-1}=P_{N-1}\left|
N\right\rangle \left\langle N\right| =0$ then
\begin{equation}
P_{N}P_{N-1}=P_{N-1}P_{N}=P_{N-1}^{2}=P_{N-1}.  \label{eq:P_N*P_(N-1)}
\end{equation}

\section{The Rayleigh-Ritz variational method}

\label{sec:RR}

If $S$ is the vector space on which $H$ operates then we can define the
projected subspace $S_{N}=P_{N}S$ and the projection of $H$ on $S_{N}$\cite%
{P68,SO96,CDL77}
\begin{equation}
H_{N}=P_{N}HP_{N}.  \label{eq:H_N}
\end{equation}
The eigenvectors $\left| N,j\right\rangle $ of $H_{N}$,
\begin{eqnarray}
H_{N}\left| N,j\right\rangle &=&E_{j}^{(N)}\left| N,j\right\rangle
,\;j=1,2,\ldots ,N,  \nonumber \\
E_{1}^{(N)} &\leq &E_{2}^{(N)}\leq \ldots \leq E_{N}^{(N)},
\label{eq:eigenvect_H_N}
\end{eqnarray}
belong to $S_{N}$
\begin{equation}
\left| N,j\right\rangle =\sum_{i=1}^{N}c_{ij}^{(N)}\left| i\right\rangle .
\label{eq:|N,j>}
\end{equation}
These eigenvectors can be constructed orthonormal: $\left\langle N,i\right.
\left| N,j\right\rangle =\delta _{ij}$. Obviously,
\begin{equation}
P_{N}\left| N-1,j\right\rangle =P_{N-1}\left| N-1,j\right\rangle =\left|
N-1,j\right\rangle .  \label{eq:P_N|N-1,j>}
\end{equation}

It is convenient to define the linear combination
\begin{equation}
\varphi =\sum_{j=1}^{K}a_{j}\left| N-1,j\right\rangle ,\;1\leq K\leq N-1,
\label{eq:varphi}
\end{equation}
and choose the coefficients $a_{j}$ so that
\begin{equation}
\left\langle N,i\right. \left| \varphi \right\rangle
=\sum_{j=1}^{K}a_{j}\left\langle N,i\right. \left| N-1,j\right\rangle
=0,\;i=1,2,\ldots ,K-1.  \label{eq:a_j_cond}
\end{equation}
It is clear that this set of $K-1$ equations with $K$ unknowns $a_{j}$ will
always have nontrivial solutions. According to equation (\ref{eq:UB_E_k}) we
have
\begin{equation}
\frac{\left\langle \varphi \right| H_{N}\left| \varphi \right\rangle }{%
\left\langle \varphi \right. \left| \varphi \right\rangle }\geq E_{K}^{(N)}.
\label{eq:UB_E_K_H_N}
\end{equation}
On the other hand, since $P_{N}\varphi =P_{N-1}\varphi =\varphi $ it follows
that
\begin{eqnarray}
\left\langle \varphi \right| H_{N}\left| \varphi \right\rangle
&=&\left\langle \varphi \right| P_{N-1}H_{N}P_{N-1}\left| \varphi
\right\rangle  \nonumber \\
&=&\left\langle \varphi \right| H_{N-1}\left| \varphi \right\rangle
=\sum_{j=1}^{K}\left| a_{j}\right| ^{2}E_{j}^{(N-1)}\leq
E_{K}^{(N-1)}\left\langle \varphi \right. \left| \varphi \right\rangle ,
\end{eqnarray}
or
\begin{equation}
\frac{\left\langle \varphi \right| H_{N}\left| \varphi \right\rangle }{%
\left\langle \varphi \right. \left| \varphi \right\rangle }\leq
E_{K}^{(N-1)}.  \label{eq:LB_E_K^(N-1)}
\end{equation}
It follows from equations (\ref{eq:UB_E_K_H_N}) and (\ref{eq:LB_E_K^(N-1)})
that $E_{K}^{(N)}\leq E_{K}^{(N-1)}$.

We next choose another linear combination
\begin{equation}
\psi =\sum_{j=1}^{K}b_{j}\left\vert N,j\right\rangle ,\;1\leq K\leq N-1,
\label{eq:psi}
\end{equation}%
and require that the coefficients $b_{j}$ satisfy
\begin{equation}
\left\langle \psi _{i}\right. \left\vert \psi \right\rangle
=\sum_{j=1}^{K}b_{j}\left\langle \psi _{i}\right. \left\vert
N,j\right\rangle =0,\;i=1,2,\ldots ,K-1.  \label{eq:b_j_cond}
\end{equation}%
Then, according to equation (\ref{eq:UB_E_k}) we have
\begin{equation}
\frac{\left\langle \psi \right\vert H\left\vert \psi \right\rangle }{%
\left\langle \psi \right. \left\vert \psi \right\rangle }\geq E_{K}.
\label{eq:UB_E_K_2}
\end{equation}%
On the other hand, it follows from
\begin{eqnarray}
\left\langle \psi \right\vert H\left\vert \psi \right\rangle
&=&\left\langle \psi \right\vert P_{N}HP_{N}\left\vert \psi \right\rangle
=\left\langle \psi \right\vert H_{N}\left\vert \psi \right\rangle
=\sum_{j=1}^{K}\left\vert b_{j}\right\vert ^{2}E_{j}^{(N)}  \nonumber \\
&\leq &E_{K}^{(N)}\left\langle \psi \right. \left\vert \psi \right\rangle ,
\end{eqnarray}%
that
\begin{equation}
\frac{\left\langle \psi \right\vert H\left\vert \psi \right\rangle }{%
\left\langle \psi \right. \left\vert \psi \right\rangle }\leq E_{K}^{(N)}.
\label{eq:LB_E_K^(N)}
\end{equation}%
It follows from equations (\ref{eq:UB_E_K_2}) and (\ref{eq:LB_E_K^(N)}) that
$E_{K}\leq E_{K}^{(N)}$. We thus derive the main property of the RRVM:
\begin{equation}
E_{K}^{(N-1)}\geq E_{K}^{(N)}\geq E_{K},\;1\leq K\leq N-1.  \label{eq:UB_E's}
\end{equation}%
It is important to realize that it is not necessary to know the eigenvectors
$\psi _{i}$ of $H$ explicitly in order to derive equation (\ref{eq:UB_E_K_2}%
). We merely assume the existence of such eigenvectors and prove that the
trial function $\psi $ already exists. From a practical point of view, the
upper bounds in equation (\ref{eq:UB_E's}) are given by the eigenvectors of $%
H_{N}$ that we can easily obtain as argued below.

Any eigenvector $\varphi $ of $H_{N}$ can be written as
\begin{equation}
\varphi =\sum_{j=1}^{N}c_{j}\left| j\right\rangle .
\label{eq:varphi_eigenvector}
\end{equation}
Therefore, it follows from $\left( H_{N}-E\right) \varphi =0$ that
\begin{eqnarray}
\left\langle k\right| \left. \left( H_{N}-E\right) \varphi \right\rangle
&=&\left\langle k\right| \left. \left( P_{N}HP_{N}-E\right) \varphi
\right\rangle =\left\langle k\right| \left. \left( H-E\right) \varphi
\right\rangle  \nonumber \\
&=&\sum_{j=1}^{N}\left( H_{k,j}-E\delta _{kj}\right) c_{j}=0,\;k=1,2,\ldots
,N,  \nonumber \\
H_{k,j} &=&\left\langle k\right| H\left| j\right\rangle .  \label{eq:secular}
\end{eqnarray}
There are nontrivial solutions to this equation if $E$ is a root of \cite%
{P68,SO96}
\begin{equation}
\left| \mathbf{H}_{N}-E\mathbf{I}\right| =0,  \label{eq:secular_det}
\end{equation}
where $\mathbf{H}_{N}$ is the $N\times N$ matrix with elements $H_{ij}$, $%
i,j=1,2,\ldots ,N$, and $\mathbf{I}$ is the $N\times N$ identity matrix. We
conclude that the roots $E_{n}^{(N)}$, $n=1,2,\ldots ,N$, of the secular
determinant (or characteristic polynomial) (\ref{eq:secular_det})\cite%
{P68,SO96} satisfy the inequalities (\ref{eq:UB_E's}) for $N=2,3,\ldots $.
In other words, the solutions of the RRVM, given by equations (\ref%
{eq:secular}) and (\ref{eq:secular_det}), approach the corresponding
eigenvalues of $H$ from above. For a given eigenvalue $E_{n}^{(N)}$ of $%
H_{N} $ we obtain the coefficients $c_{jn}^{(N)}$ from equation (\ref%
{eq:secular}) and then the eigenvectors $\left| N,j\right\rangle $ shown in
equation (\ref{eq:|N,j>}) (after suitable normalization)\cite{P68,SO96}.

More details about the convergence of the RRVM in the case of the
Hamiltonian operators that appear in molecular physics and quantum chemistry
were given by Klahn and Bingel\cite{KB77} some time ago.

\section{Example}

\label{sec:example}

As an example we consider a simple model given by a particle of effective
mass $m^{*}$ and charge $e$ in an infinite square well of length $L$ under
the effect of an electric field of intensity $F$\cite{BMCE83}. The
Hamiltonian operator reads
\begin{equation}
H=-\frac{\hbar ^{2}}{2m^{*}}\frac{d^{2}}{dz^{2}}+|e|Fz,  \label{eq:H_PB}
\end{equation}
and its eigenfunctions $\psi _{n}(z)$ satisfy $\psi _{n}(0)=\psi _{n}(L)=0$.

In order to facilitate the calculation it is convenient to convert the Shcr%
\"{o}dinger equation into a dimensionless eigenvalue equation. To this end,
we define the dimensionless coordinate $\tilde{z}=z/L$ and Hamiltonian (see%
\cite{F20} for more details and examples)
\begin{equation}
\tilde{H}=\frac{m^{*}L^{2}}{\hbar ^{2}}H=-\frac{1}{2}\frac{d^{2}}{d\tilde{z}%
^{2}}+\lambda \tilde{z},\;\lambda =\frac{m^{*}|e|FL^{3}}{\hbar ^{2}}.
\label{eq:H_PB_dim}
\end{equation}
The eigenvalues $E_{n}$ and $\tilde{E}_{n}$, $n=1,2,\ldots $, of $H$ and $%
\tilde{H}$, respectively, are related by $E_{n}(m^{*},e,F,L)=\hbar ^{2}%
\tilde{E}_{n}(\lambda )/\left( m^{*}L^{2}\right) $. Notice that the
dimensionless constant $\lambda $ is the most relevant parameter of the model%
$.$ The eigenfunctions $\tilde{\psi}_{n}$ of $\tilde{H}$ satisfy the
boundary conditions $\tilde{\psi}_{n}(0)=\tilde{\psi}_{n}(1)=0$.

A suitable orthonormal basis set of functions for the application of the
RRRVM is
\begin{equation}
u_{j}(\tilde{z})=\sqrt{2}\sin (j\pi \tilde{z}),\;j=1,2,\ldots ,
\label{eq:u_j}
\end{equation}
that satisfy the required boundary conditions. The matrix elements of $%
\tilde{H}$ can be obtained analytically
\begin{equation}
\tilde{H}_{i,j}=\left\{
\begin{array}{c}
\frac{\pi ^{2}i^{2}}{2}+\frac{\lambda }{2},\;i=j \\
\frac{4ij\left[ \left( -1\right) ^{i+j}-1\right] }{\pi ^{2}\left(
i^{2}-j^{2}\right) ^{2}}\lambda ,\;i\neq j%
\end{array}
\right. ,\;i,j=1,2,\ldots .  \label{eq:H_PB_(i,j)}
\end{equation}
Table~\ref{tab:conv_RR} shows the rate of convergence of the first four
eigenvalues $\tilde{E}_{n}$ for $\lambda =1$. It is clear that the sequence $%
\tilde{E}_{n}^{(N)}$, $N=2,3,\ldots $, converges from above as proved in
section~\ref{sec:RR}.

The eigenvalue equation
\begin{equation}
\psi ^{\prime \prime }\left( \tilde{z}\right) +2\left( \tilde{E}-\lambda
\tilde{z}\right) \psi (\tilde{z})=0,  \label{eq:eig_eq_model}
\end{equation}
can be transformed into the Airy equation\cite{AS72}
\begin{equation}
y^{\prime \prime }(q)-qy(q)=0,  \label{eq:Airy}
\end{equation}
by means of the change of variables
\begin{equation}
q=(2\lambda )^{1/3}\left( z-\frac{\tilde{E}}{\lambda }\right) .
\label{eq:q(z)}
\end{equation}
The solution to equation (\ref{eq:eig_eq_model}) can therefore be written as
\begin{equation}
\tilde{\psi}\left( \tilde{z}\right) =c_{1}Ai(q)+c_{2}Bi(q),
\label{eq:psi(z)}
\end{equation}
where $A_{i}(q)$ and $B_{i}(q)$ are the Airy functions of the first kind\cite%
{AS72}. The boundary conditions $\tilde{\psi}(0)=\tilde{\psi}(1)=0$ lead to
a linear system of two equations with two unknowns, $c_{1}$ and $c_{2}$,
with nontrivial solutions only for those values of $\tilde{E}=\tilde{E}_{n}$
that are roots of the equation
\begin{eqnarray}
&&Ai\left( q_{L}\right) Bi\left( q_{R}\right) -Ai\left( q_{R}\right)
Bi\left( q_{L}\right) =0,  \nonumber \\
&&q_{L}=-\left( 2\lambda \right) ^{1/3}\frac{\tilde{E}}{\lambda }%
,\;q_{R}=\left( 2\lambda \right) ^{1/3}\left( 1-\frac{\tilde{E}}{\lambda }%
\right) .  \label{eq:quant_cond}
\end{eqnarray}
One can easily verify that the \textit{exact }eigenvalues obtained in this
way agree with those provided by the RRVM.

\section{Conclusions}

\label{sec:conclusions}

In this paper we have derived an important property of the RRVM in a way
that may be suitable for an advanced course on quantum chemistry or quantum
mechanics. Present approach illustrates the utility of a widely used kind of
projection operators\cite{P68,SO96}. It is worth noticing that the arguments
based on the vectors (\ref{eq:varphi}) and (\ref{eq:psi}) are almost
identical, the only difference being that we compare the eigenvalues of $%
H_{N}$ and $H_{N-1}$ with the former and the eigenvalues of $H$ and $H_{N}$
with the latter. We think that our strategy, which generalizes the exercise
proposed by Szabo and Ostlund\cite{SO96}, is simpler than the one developed
by MacDonald\cite{M33} several years ago. As stated above, Klahn and Bingel%
\cite{KB77} discussed the convergence of the RRVM in the case of the
Hamiltonian operators that commonly appear in molecular physics and quantum
chemistry. Here, we illustrated the performance of the method by means of a
simple one-dimensional model suitable for an advanced course on quantum
mechanics or quantum chemistry.

\begin{table}
\caption{Convergence of the RRVM for the first
eigenvalues of the dimensionless Hamiltonian (\ref{eq:H_PB_dim})
with $\lambda=1$.}

\label{tab:conv_RR}

\begin{tabular}{D{.}{.}{3}D{.}{.}{11}D{.}{.}{11}D{.}{.}{11}D{.}{.}{11}}
 \multicolumn{1}{l}{$N$}&\multicolumn{1}{c}{$\tilde{E}_1$}&
\multicolumn{1}{c}{$\tilde{E}_2$} &
\multicolumn{1}{c}{$\tilde{E}_3$} & \multicolumn{1}{c}{$\tilde{E}_4$} \\
\hline
2&   5.432610908 & 20.24140009 &\multicolumn{1}{c}{-} & \multicolumn{1}{c}{-}  \\
4&   5.432607957 & 20.23986646 & 44.91361286 & 79.45797872 \\
6&   5.432607865& 20.23986320& 44.91360984& 79.45707684  \\
8&   5.432607857& 20.23986306& 44.91360969& 79.45707417  \\
10&  5.432607855& 20.23986304& 44.91360967& 79.45707402  \\
12&  5.432607855& 20.23986304& 44.91360966& 79.45707400  \\
14&  \multicolumn{1}{c}{-}& \multicolumn{1}{c}{-}& 44.91360966& 79.45707400  \\

\end{tabular}

\end{table}

\end{document}